\documentclass[twocolumn]{aastex6}
\usepackage{epstopdf}
\usepackage{amsmath,array,graphicx}

\usepackage{array,graphicx,tabularx,threeparttable, capt-of}
\usepackage{booktabs}
\usepackage{pifont}
\usepackage{multirow}
\usepackage{natbib}
\usepackage{color}

\fullcollaborationName{The Friends of AASTeX Collaboration}

\begin{document}


\title{Nature of stochastic ion heating in the solar wind:\\
testing the dependence on plasma beta and turbulence amplitude}


\author{Daniel Vech\altaffilmark{1}, Kristopher G. Klein\altaffilmark{1,2} and Justin C. Kasper\altaffilmark{1},}

\affil{$^{1}$Climate and Space Sciences and Engineering, University of Michigan, Ann Arbor, MI 48109, USA, $^{2}$Lunar and Planetary Laboratory, University of Arizona, Tucson, AZ 85719, USA; dvech@umich.edu}
\begin{abstract}
The solar wind undergoes significant heating as it propagates away
from the Sun; the exact mechanisms responsible for this heating are
not yet fully understood. We present for the first time a statistical
test for one of the proposed mechanisms, stochastic ion heating. We use
the amplitude of magnetic field fluctuations near the
proton gyroscale as a proxy for the ratio of gyroscale velocity
fluctuations to perpendicular (with respect to the magnetic field) proton thermal speed, defined as $\epsilon_p$. Enhanced proton temperatures are observed when $\epsilon_p$ is larger than a critical value ($\sim 0.019 - 0.025$). This enhancement
strongly depends on the proton plasma beta ($\beta_{||p}$); when
$\beta_{||p} \ll 1$ only the perpendicular proton temperature $T_{\perp}$ increases, while for $\beta_{||p} \sim 1$ increased parallel and perpendicular proton temperatures are both observed. For
$\epsilon_p$ smaller than the critical value and $\beta_{||p} \ll 1$
no enhancement of $T_p$ is observed while for $\beta_{||p} \sim 1$ minor
increases in $T_{\parallel}$ are measured. The observed change of proton
temperatures across a critical threshold for velocity fluctuations is in
agreement with the stochastic ion heating model of \citet{chandran2010perpendicular}. We find that $\epsilon_p > \epsilon_{\rm crit}$ in 76\% of the studied periods implying that stochastic heating may operate most of the time in the solar wind at 1 AU.

\end{abstract}

\keywords{plasmas --- turbulence --- solar wind --- waves}



\section{Introduction}\label{sec:intro}

The solar wind is a hot, tenuous plasma propagating away from Sun's
surface. The radial expansion of the solar wind is highly
non-adiabatic with the proton temperature cooling significantly slower
than a spherically expanding ideal gas \citep[e.g.][]{wolfe1966observations, hundhausen1970solar}. The radial dependence of proton temperature $T_p$ as a function of the
heliocentric distance $r$ is measured on the average as $r^{-0.74}$ compared to
$r^{-4/3}$ corresponding to adiabatic expansion
\citep[]{hellinger2011heating}. This slow decay of the temperature is consistent
with the solar wind undergoing significant heating. Identifying the
physical mechanisms responsible for this heating and quantifying their
contribution as a function of plasma and solar wind parameters is
fundamentally important to describing the solar corona and solar wind
and to characterizing heating in plasma systems more
generally. Several mechanisms have been proposed to heat the solar
wind as it expands, including cyclotron damping
\citep[]{cranmer2000ion}, magnetic reconnection
\citep[]{drake2009ion, osman2012kinetic, osman2014magnetic, greco2016complex, mistry2017statistical}, Landau damping \citep[]{leamon1999dissipation,
  cranmer2007self, gary2004kinetic} and stochastic heating
\citep[]{mcchesney1987observation, johnson2001stochastic,
  chaston2004auroral, voitenko2004cross, van2014alfven}.

This Letter focuses on stochastic ion heating: such heating occurs
when the motion of ions becomes chaotic as the amplitude of
electromagnetic field fluctuations at scales comparable to the ion
gyroscale exceed a critical value. Under these conditions, the
magnetic moment of ions is not conserved, allowing diffusion in energy perpendicular to the magnetic field and leading to perpendicular heating of the ions. Stochastic heating may have a
significant contribution to the ion heating in coronal holes and the
solar wind, however its importance relative to other mechanisms is an
open question. Coronagraph measurements have shown that minor ions
such as O$^{+5}$ originating from coronal holes have significantly
larger perpendicular temperature $T_{\perp}$ (with respect to the
magnetic field) than parallel $T_{||}$ \citep[]{kohl1998uvcs,
  antonucci2000fast}. In situ observations of the fast solar wind
frequently find similar proton temperature anisotropy of
$T_{\perp}/T_{||}>1$ \citep[e.g][]{marsch2004temperature,
  hellinger2006solar}.

\citet{chandran2010perpendicular} modeled ion stochastic heating by
low frequency ($\omega<\Omega_p$, where $\omega$ and $\Omega_p$
denote the wave and proton cyclotron frequencies, respectively)
Alfv\'en (AW) and kinetic Alfv\'en waves (KAW). They proposed that the
heating rate of this mechanism is very sensitive to the amplitude of
the turbulent velocity fluctuations, which they characterized by the
dimensionless parameter $\epsilon_i=\delta v_{\rho}/v_{\perp}$
where $\delta v_{\rho}$ denotes the amplitude of root-mean-square
velocity fluctuations at scales comparable to the ion gyroscale while
$v_{\perp}$ is the ion's thermal speed perpendicular to the background magnetic field. When the velocity fluctuations are smaller than some
critical value, $\epsilon_i \ll \epsilon_{\rm crit}$, the magnetic
moment of the ions is conserved and any stochastic heating is
suppressed. When $\epsilon_i \gtrsim \epsilon_{\rm crit}$ magnetic
moment conservation is violated,leading to energy diffusion perpendicular to the magnetic field and an increase in $T_\perp$. In test particle
simulations from \citet{chandran2010perpendicular} $\epsilon_{\rm
  crit}$ was reported as 0.19. \citet{chandran2010perpendicular} predicted that
depending on the values of $\beta_{||p}$ (ratio of parallel thermal
pressure to magnetic pressure; $n_p k_BT_{||p}/(B_0^2/2\mu_0$), where
$n_p$ denotes the proton density, $k_B$ is the Boltzmann constant,
$T_{||p}$ is the parallel proton temperature, $B_0$ is the magnitude of
the magnetic field, and $\mu_0$ is the permeability of free space) the
following proton heating behaviors are expected under the assumption of low-frequency, KAW-like turbulence:

\begin{enumerate}
\item If $\beta_{\parallel p} \ll 1$ and $\epsilon_p \ll \epsilon_{\rm
  crit}$, electrons absorb the vast majority of the cascade power and
  proton heating is negligible as the Landau resonance condition
  ($\omega - k_{\parallel}v_{\parallel}=0$, where $k_{\parallel}$ and
  $v_{\parallel}$ denote the parallel wavenumber and particle velocity
  along the magnetic field direction, respectively) is not satisfied
  for protons.
  \citep[]{quataert1998particle, gruzinov1998radiative}.
\item If $\beta_{\parallel p} \ll 1$ and $\epsilon_p \gtrsim
  \epsilon_{\rm crit}$, stochastic ion heating operates and AW/KAW
  turbulence causes both electron and perpendicular proton heating
  while the parallel proton heating is negligible.
\item If $\beta_{\parallel p} \sim 1$ and $\epsilon_p \ll
  \epsilon_{\rm crit}$, electron and parallel proton heating occurs
  due to Landau damping and transit-time damping of KAWs. Stochastic
  heating is suppressed producing no increase in the perpendicular
  proton temperature.
\item If $\beta_{\parallel p} \sim 1$ and $\epsilon_p \gtrsim
  \epsilon_{\rm crit}$, stochastic heating operates and the electrons
  and protons both receive significant fractions of the cascade power,
  with similar energy transferred to both perpendicular and parallel
  proton temperatures.
\end{enumerate}

 \citet[]{bourouaine2013observational}
tested the predictions of \citet{chandran2010perpendicular} studying 3
days of Helios-2 measurements with radial distances ranging from 0.29-0.64 AU, focusing on fast solar wind with low plasma beta ($\beta <
$ 0.3). Stochastic heating by low-frequency AW/KAW
turbulence was consistent with the observed perpendicular temperature
in the solar wind for the three selected
intervals. \citet[]{xia2013perpendicular} performed further tests of the
model of \citet{chandran2010perpendicular}, describing test particles interacting with strong reduced
magnetohydrodynamic (RMHD) turbulence. RMHD was found to be much more
effective at stochastic heating than randomly phased waves used in
previous studies. They suggested that stochastic heating can occur not
only under $\beta \ll 1$ conditions of the solar corona but also when
$\beta \sim 1$, typically occurring at 1
AU. \citet[]{klein2016evolution} modeled the evolution of proton
distributions due to stochastic heating in the range of 4 to 30 solar radii, finding that the proton
distributions developed non-Gaussian structures characterized with a flat core and steep tail.

Despite these works, the role of stochastic heating in the solar wind
is not yet fully understood. In particular no statistical study using
solar wind observations has been carried out to test the effect of
low-frequency AW/KAWs on proton heating. In this Letter, we present
the analysis of 13 years of Wind data to investigate scalar proton and
electron temperatures and proton temperature anisotropy as a function
of $\beta$ and turbulence amplitude. We expect and find positive correlation between the turbulence amplitude and $T_p$, in agreement with previous studies \citep[e.g.][]{grappin1990origin, cranmer2005generation, wu2013karman, matthaeus2016turbulence, hughes2017particle, hughes2017kinetic}, 
suggesting that the damping mechanism requires a higher amplitude of the fluctuations in order to operate. We explicitly compare the observed features
to the expected behavior of stochastic heating and find the proton temperature depends on a critical turbulence amplitude in agreement with
the predictions of \citet[]{chandran2010perpendicular}.

\section{Method} \label{sec:results}

\subsection{Background} \label{subsec:theory}

The goal of the data analysis was to organize temperature measurements
as a function of ($\epsilon_p$, $\beta_{||p}$) and compare the
temperature dependence of these parameters with the predictions of
stochastic ion heating. Following the methodology of
\citet[]{bourouaine2013observational}, we define the velocity
fluctuations at the proton gyroscale, $\delta v_p$, as
\begin{equation}
\delta v_p = \sigma v_A \delta B_p / B_0
\label{eqn:dvp}
\end{equation}
where $\sigma=1.19$ is a dimensionless constant arising from the KAW
dispersion relation, $v_A=B_0/ \sqrt{\mu_0 \rho}$ denotes the Alfv\'en
speed, $\rho$ is the
mass density of the solar wind,
and $\delta B_p$ is the amplitude of magnetic field fluctuations at proton gyroscale.  This approximation is necessary as observations
of velocity fluctuations have not yet been made at sufficiently high
cadences to resolve $\delta v_p$ except in unusual solar wind conditions. 
In Equation~\ref{eqn:dvp}, $\delta B_p$ is obtained by
evaluating
\begin{equation}
\delta B_{p} = \frac{\pi}{C_{\gamma}} \left[
  \int_{e^{-0.5}f_{\rho}}^{e^{0.5}f_{\rho}} P_f(f) df \right]^{1/2}
\label{eqn:dbp}
\end{equation}
where $P_f(f)$ denotes the observed turbulent spectrum of magnetic
fluctuations, $f_{\rho}$ is the frequency corresponding to the proton
gyroradius defined as $V_{SW} \sin(\Theta_{VB})/2\pi \rho_p$, where
$V_{SW}$ is the speed of the solar wind, $\Theta_{VB}$ is the angle
between the solar wind velocity vector and the magnetic field and
$\rho_p$ is the proton gyroradius. The dimensionless parameter
$C_{\gamma}$ arises from the integral $\int_0^{\pi/2}(\cos
\phi)^{\gamma - 1} d \phi$, and is defined as
\begin{equation}
C_{\gamma} = \frac{ \sqrt{\pi} \Gamma \left[ \frac{\gamma}{2} \right]}
{2 \Gamma \left[ \frac{\gamma + 1}{2} \right]}
\label{eqn:C}
\end{equation}
where $\gamma$ is the absolute value of the spectral index of the
turbulent spectrum within the integration limits. An extensive
discussion of using magnetic fluctuation frequency spectra as a proxy
for gyroscale velocity fluctuations can be found in Appendices A and B
of \citet[]{bourouaine2013observational}.

Equation~\ref{eqn:dbp} assumes that $C_{\gamma}$ is constant within
the integration limits, meaning that the spectral index $\gamma$ does
not change in the given frequency range. This requirement is violated
when the break frequency $f_b$ is within the integration limits,
$e^{-0.5f_{\rho}} < f_b <
e^{0.5f_{\rho}}$. \citet[]{bourouaine2013observational} restricted
their analysis to 3 intervals when the integration limits were above
the break of the turbulent spectrum, $f_b < e^{-0.5f_{\rho}}$. As we
aim to use a statistical approach, we employ the following
approximation when $f_b$ is within the integration limits of
Equation~\ref{eqn:dbp}. We replace $C_\gamma$ in
Equation~\ref{eqn:dbp} with $\bar{C_{\gamma}}$, the weighted average
of $C_{\gamma}$ below and above $f_b$:

\begin{equation}
\bar{C}_{\gamma} = \frac{ C_{\gamma 1} \int_{e^{-0.5}f_{\rho}}^{f_b }
  P_f(f) df + C_{\gamma 2} \int_{f_b}^{e^{0.5}f_{\rho} } P_f(f) df } {
  \int_{e^{-0.5}f_{\rho}}^{f_b } P_f(f) df +
  \int_{f_b}^{e^{0.5}f_{\rho} } P_f(f) df }
\label{eqn:Cbar}
\end{equation}
where $C_{\gamma 1,2}$ is calculated separately for the spectral
indices above and below $f_b$ using Equation~\ref{eqn:C}. This
approximation is used in 31\% percent of the intervals.
Given $\delta B_p$ calculated from Equations~\ref{eqn:dbp} and \ref{eqn:Cbar}, we
calculate $\epsilon_p = \delta v_p/v_{\perp}$ using Equation~\ref{eqn:dvp}
and the perpendicular proton thermal speed $v_{\perp} = \sqrt{2 k_B T_{\perp}/m_p}$.

\subsection{Application} \label{subsec:app}

\begin{figure*}[ht!]
\figurenum{1} \plotone{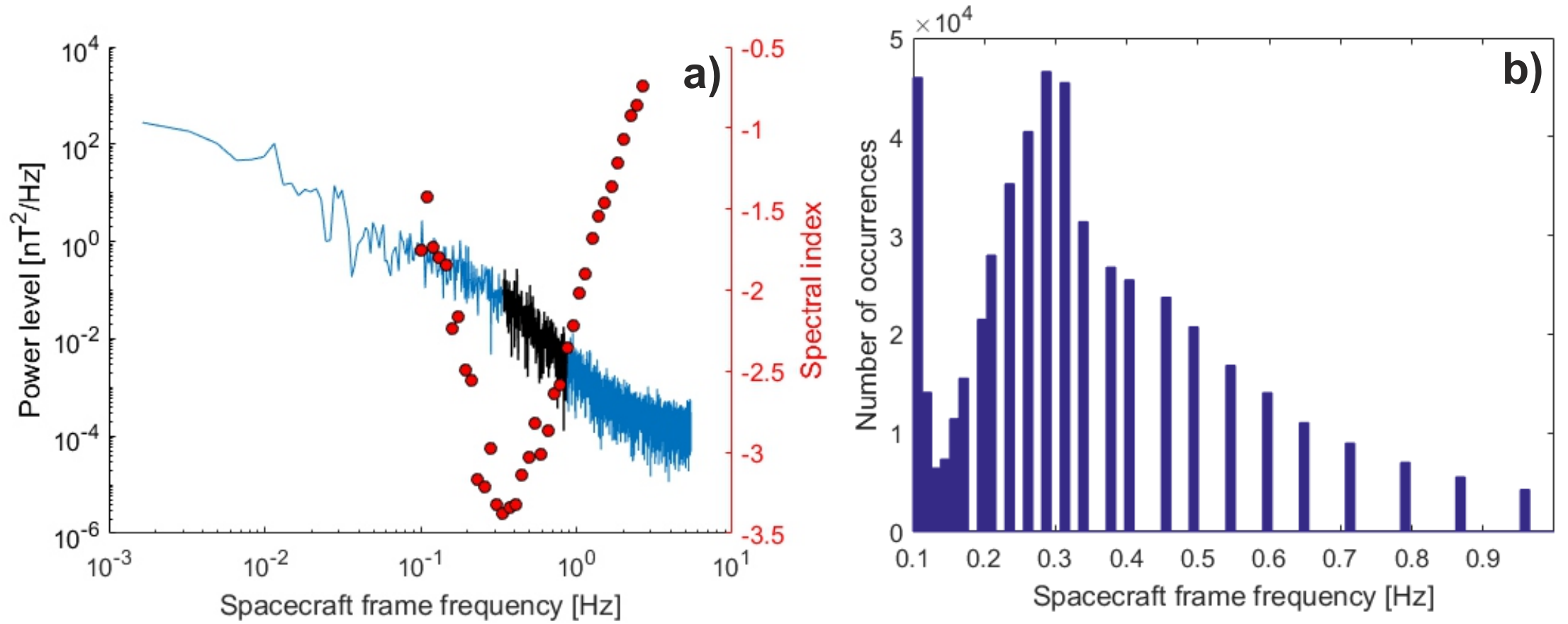}
\caption{(a) an example of the measured magnetic turbulent spectrums
  and the measured spectral indices (red marks) in the 0.1-5.17 Hz
  range; (b) histogram of the low frequency end of the measured
  dissipation ranges (black part of the spectrum in panel a) for all
  the $\sim 5.8 \cdot 10^5$ periods.  }
  \label{fig:1}
\end{figure*}

In this study, high resolution Wind magnetic field data (92 ms
cadence) \citep[]{lepping1995wind} were used together with onboard ion
moments and ion parameters (92 second cadence) from the Faraday cup
instrument \citep[]{lin1995three, ogilvie1995swe}. Data from January
2004 to December 2016 were selected to ensure Wind was in the pristine
solar wind. For the analysis, the magnetic
field and plasma data were split in 10-minute intervals. The power spectral density (PSD) of the magnetic field
components were calculated separately using Fourier transform and then
the component PSDs were added to obtain the total PSD
\citep[]{koval2013magnetic}. The time series of $\beta_{||p}$,
$T_{||}$, $T_{\perp}$ and electron temperature ($T_e$) were averaged
over the 10-minute periods. Overall, $\sim 5.8 \cdot 10^5$ turbulent
spectrums and corresponding average solar wind parameters were
computed. Due to gaps in the data, only $\sim 5.2 \cdot 10^5$ average
electron temperatures were obtained.

For the correct calculation of $C_{\gamma}$ and $\bar{C}_{\gamma}$
(Equations \ref{eqn:C}-\ref{eqn:Cbar}), it was necessary to estimate $f_b$,
which shows some variability ranging from 0.1-1 Hz
\citep[]{markovskii2008statistical, chen2014ion, franci2016plasma, telloni2016linking} making its
parameterization difficult. To automatically estimate this frequency,
we developed the following algorithm: starting from 0.1 Hz until 5.17
Hz a grid of 43 logarithmically spaced frequencies was generated. For
each PSD, 33 linear fits were made in the frequency range between the
i$^{th}$ and i$+$10$^{th}$ element of the grid. From the ensemble of
fits, the steepest spectral index and the corresponding frequency
range were selected. The average and standard deviation of the
measured spectra indices are $-2.99 \pm 0.65$, in very good agreement
with previous studies on the dissipation range from
\citet[]{leamon1998observational} and \citet[]{smith2006dependence}.

The frequency range corresponding to the steepest part
of the spectrum can be used very effectively to estimate $f_b$, which
is shown in Figure \ref{fig:1}a and b. A typical magnetic field turbulent
spectrum is presented in Figure \ref{fig:1}a. The steepest part of the spectrum
(corresponding to the dissipation range) was detected automatically
with our algorithm and is marked with a black line. The red circles
denote the 33 spectral indices in the range of $0.1-5.17$ Hz. In
Figure \ref{fig:1}b, we investigate how well this method could be used as a
proxy for $f_b$ on a statistical basis. A histogram of the low
frequency end of the dissipation range is illustrated, based on all
the available $\sim 5.8 \cdot 10^5$ data points. The distribution has
a peak at around 0.3 Hz, decaying rapidly toward larger frequencies
with only 0.8\% of the distribution having $f_b$ larger than 1 Hz (not
shown). There is a secondary peak between
0.1-0.126 Hz, the majority of these spectrums did not display a
well-defined high frequency break because the spectrum flattened
immediately after the inertial range due to reaching the noise floor
\citep[]{koval2013magnetic}. This occurs when
the amplitude of the inertial scale magnetic field fluctuations are
very small, reducing the power level of the spectrum. The measurements
in the range of 0.1-0.126 Hz ($\sim 11\%$ of the overall spectrums)
shown in Figure \ref{fig:1}b were excluded from the study. The remaining
frequencies had a median value of 0.3 Hz, which is in very good
agreement with the study of \citet[]{markovskii2008statistical} who
manually inspected 454 magnetic turbulent spectrums and found that the
median of $f_b$ was approximately 0.3 Hz. They also found that $f_b$
was larger than 1 Hz in 2.1\% of the cases and it was lower than 0.1
Hz in 4.3\% of the cases.

To accurately evaluate the integral in Equation \ref{eqn:dbp}, the
unphysical flattening of the high frequency part of the spectrum must
be considered. In the cases when the high frequency end of the integration limit
$e^{0.5}f_{\rho}$ was outside the dissipation range (black
region in Figure ~\ref{fig:1}a) linear extrapolation was used to estimate the
power of the turbulent spectrum at $f_{\rho}$
\citep[]{bourouaine2013observational}. When $f_b$ was within the integration
limits of Equation~\ref{eqn:dbp}, linear fits were used in the ranges of
[$e^{-0.5}f_{\rho}$, $f_b$] and [$f_b$, $e^{0.5}f_{\rho}$] to
calculate $\gamma_{1,2}$ and $\bar{C}_{\gamma}$. The integration of
Equation \ref{eqn:dbp} was done with trapezoid technique to obtain
$\delta B_p$. Table 1 compares the results of
\citet[]{bourouaine2013observational} in the range of $0.29-0.64 $ AU
with the median values calculated from our study.

\begin{table}
  \caption{Comparison of $\delta B_p$, $\delta v_p$ and $\epsilon_p$
    from \citet[]{bourouaine2013observational} with the values
    presented in our study}
  \begin{threeparttable}
    \begin {tabularx}{\linewidth}{c*{3}{>{\hskip0pt}X}}
      \toprule %
      &Parameter & Measurements at 0.29, 0.4 and 0.64 AU & Median value of our study at 1 AU \\ \toprule %
       & $\delta B_p$ [nT]  & 1.16; 0.70; 0.32 & 0.20 \\
       & $\delta v_p$ [km/s]  & 5.15; 4.13; 3.21 & 2.34 \\
       & $\epsilon_p$  & 0.0471; 0.0486; 0.0480 & 0.0520 \\\midrule
    \end {tabularx}
  \end{threeparttable}
\end{table}

\section{Results} \label{sec:res}

\begin{figure}
\figurenum{2}
    \centering\includegraphics[width=1\linewidth]{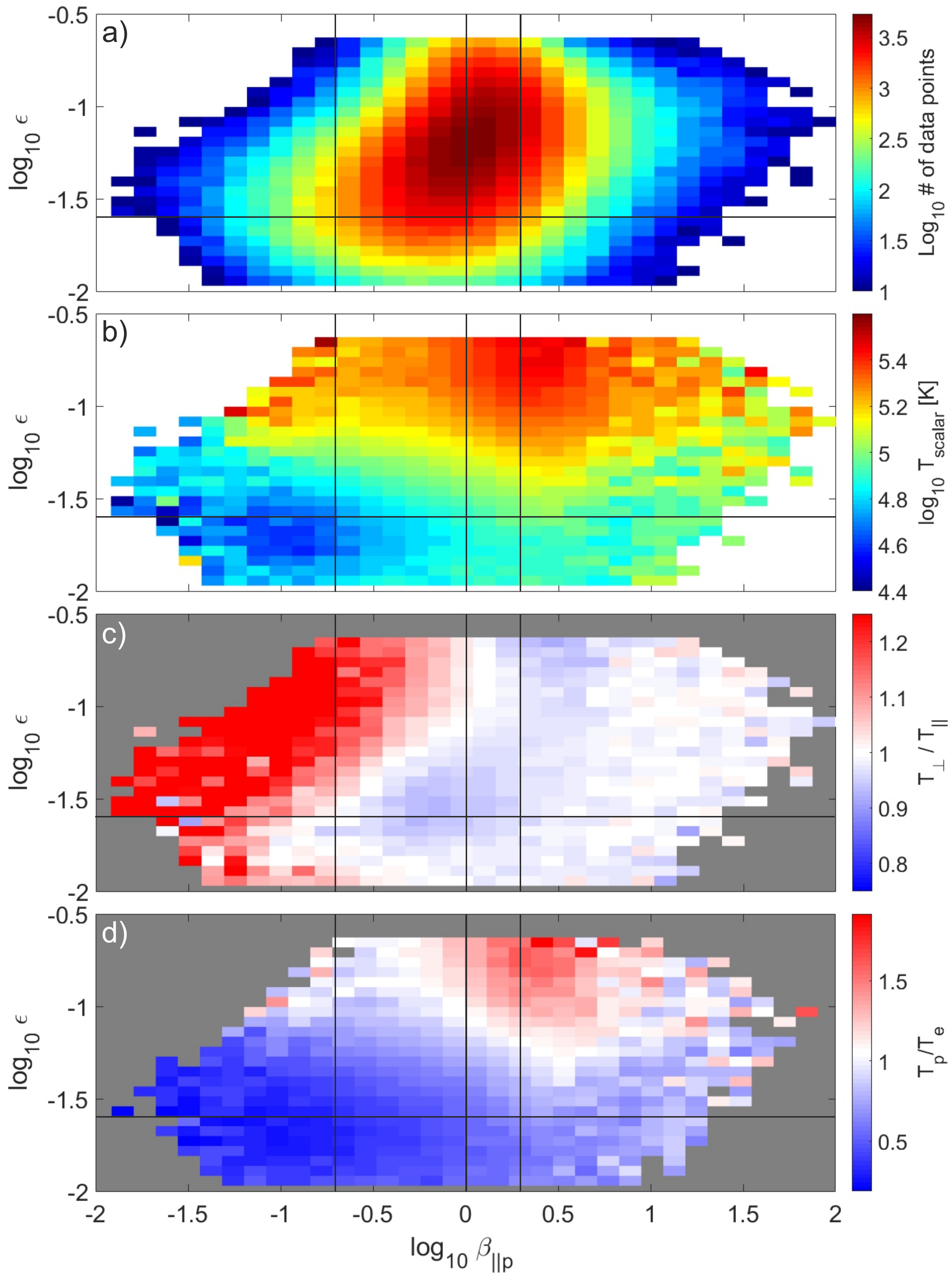}
    \caption{Data distribution (a), median values of the scalar proton temperature (b), proton temperature anisotropy (c) and proton-electron temperature ratio (d) in the ($\epsilon_p$, $\beta_{||p}$) space. Horizontal line at $\epsilon_p=10^{-1.6}$ denotes the point where stochastic heating starts operating. Vertical lines at $\beta=$0.2, 1 and 2 mark the cross sections, which are highlighted in Figure \ref{fig:3}.}
    \label{fig:2}
    \end{figure}

    \begin{figure}
    \figurenum{3}
    \centering\includegraphics[width=1\linewidth]{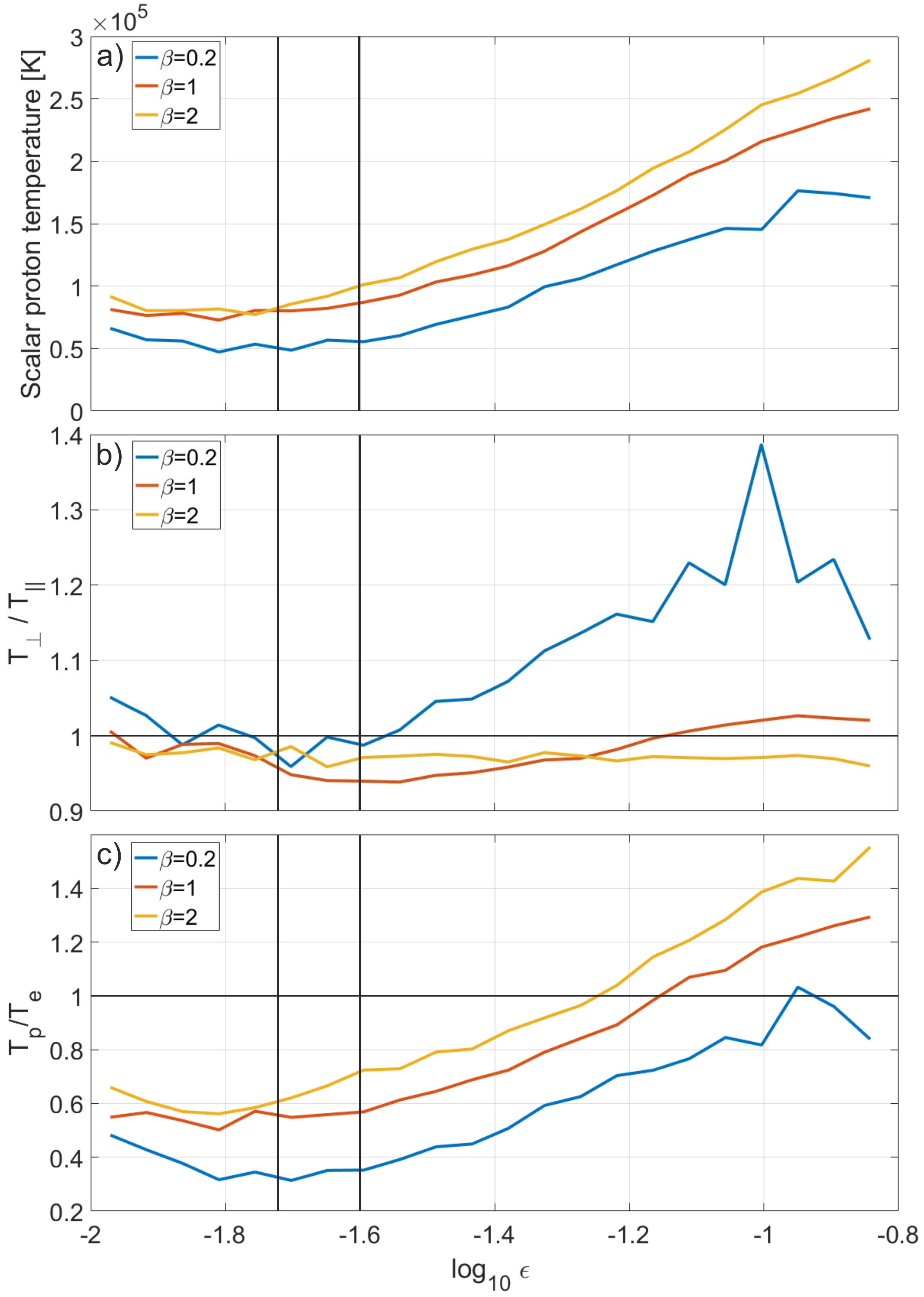}
    \caption{Cross sections of Figure \ref{fig:2}a, b and c along $\beta_{||p}=$0.2, 1 and 2, respectively.}
    \label{fig:3}
\end{figure}%

    \begin{figure}
    \figurenum{4}
    \centering\includegraphics[width=1\linewidth]{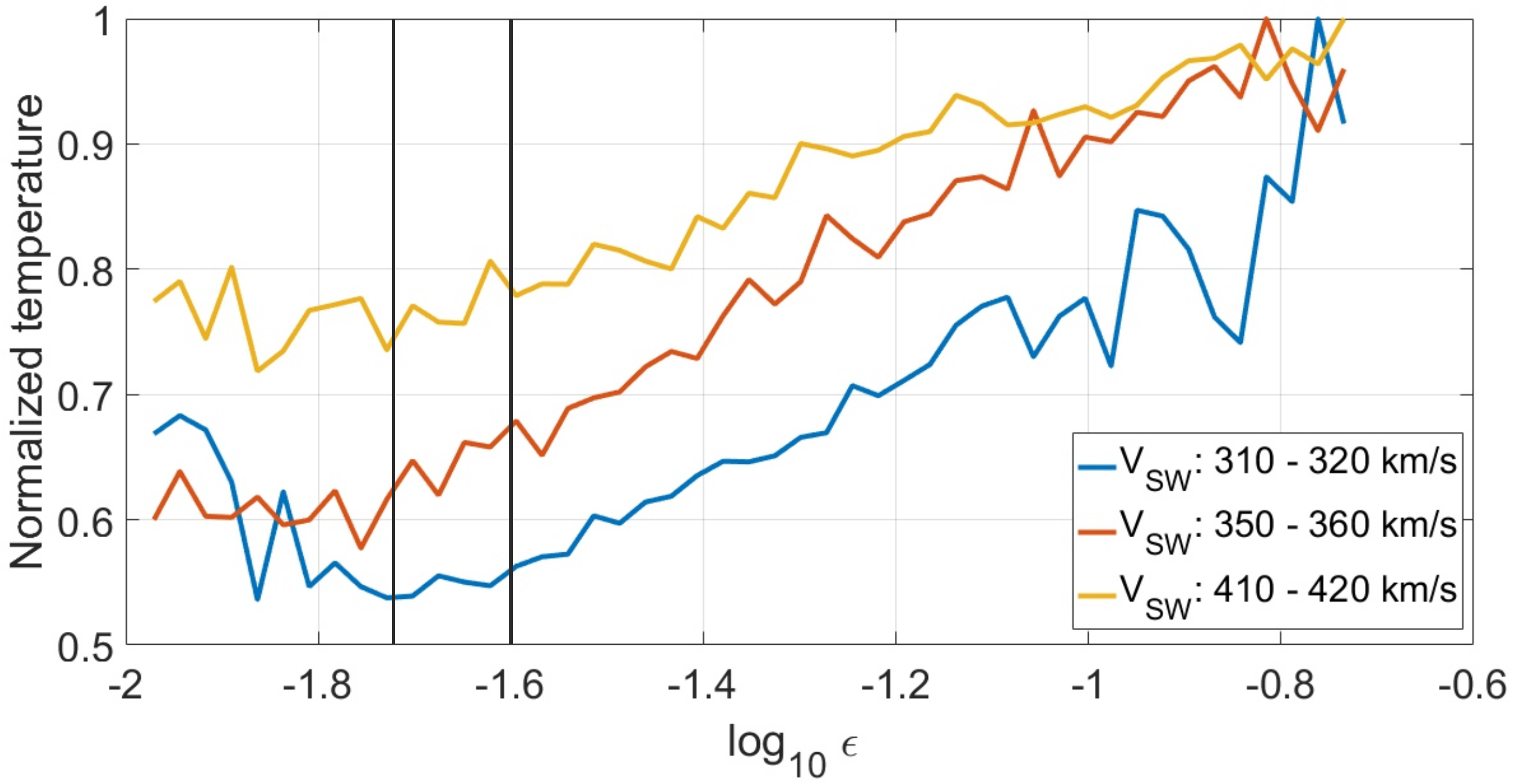}
    \caption{Cross sections of the binned $T_p$ data in the ($\epsilon_p,V_{SW}$) space along three solar wind speed intervals. Each $T_p$ line was normalized to its peak value.}
    \label{fig:4}
\end{figure}%

In order to study the proton temperature distribution, a grid with 50
x 25 equally logarithmic spaced bins was generated in the ($\epsilon_p$,
$\beta_{||p}$) space. The scalar proton temperature ($ (T_{||}+2
T_{\perp})/3$), proton temperature anisotropy ($T_{\perp}/T_{||}$) and
ratio of the scalar proton and electron temperature ($T_p/T_e$) were
binned in the defined grid. The median value of each bin was selected
and sparse bins with less than 10 data points were excluded from the
study. To avoid the possible effect of outliers, we excluded the
lowest and highest 1\% of $\epsilon_p$ values.

Figure \ref{fig:2}a shows the distribution of the data peaking at $\beta_{||p} =0.99$ and $\epsilon = 0.0520$.
The color bars in Figure \ref{fig:2}b, c and d show the binned scalar proton
temperature (on logarithmic scale), proton temperature anisotropy (on
linear scale) and proton-electron temperature ratio (on linear scale),
respectively. Cross sections of Figure \ref{fig:2}b, c and d along
$\beta_{||p}=$0.2, 1 and 2 (marked with vertical lines) as a function of $\epsilon_p$ are shown in Figure \ref{fig:3}a, b and c, respectively.
The scalar proton temperature in Figure \ref{fig:2}b shows a clear dependence on
$\epsilon_p$ and a sharp increase in the temperature can be seen at
approximately $\epsilon_p=10^{-1.6}$, marked with a black line. When
$\epsilon_p$ is smaller than $10^{-1.6}$ the temperature is around $5
\cdot 10^4$ K while for $\epsilon_p > 10^{-1.6}$ the peak temperature
is $3.1 \cdot 10^5$ K. $\epsilon_p >10^{-1.6}$ occurred in 76\% of
the $\sim 5.8\cdot10^5$ studied intervals. The $\beta_{||p}$
dependence of the scalar proton temperature is shown in Figure \ref{fig:3}a: in
all cases a sudden temperature enhancement can be seen when
$\epsilon_p$ is in the range of $10^{-1.72}$ and $10^{-1.6}$ (marked
with vertical lines).

In Figure \ref{fig:2}c, the proton temperature anisotropy increases as a
function of $\epsilon_p$ when $\beta_{||p}<1$ while no significant
systematic trend can be seen for $\beta_{||p}>1$. In Figure \ref{fig:3}b the
cross section at $\beta_{||p}>0.2$ shows some variations around
$T_{\perp}/T_{||}=1$ when $\epsilon_p < 10^{-1.6}$. For $\epsilon_p >
10^{-1.6}$ there is a significant increase in the perpendicular proton
temperature, resulting in $T_{\perp}/T_{||} \sim 1.20$. In the case of
$\beta_{||p}=1$ and $\epsilon_p < 10^{-1.6}$, $T_{\perp}/T_{||}$ shows minor preference for an enhanced parallel temperature
($T_{\perp}/T_{||} \sim 0.96$) while for $\epsilon_p > 10^{-1.6}$ $T_{\perp}/T_{||}$ approaches unity.

The $T_p/T_e$ distribution in Figure \ref{fig:2}d shows similarities to the
scalar proton temperature in Figure \ref{fig:2}b, with the ratio strongly depending
on $\epsilon_p$, having its lowest values for $\epsilon_p <
10^{-1.6}$. Similar to Figure \ref{fig:3}a and b, the cross sections in panel c
show a sudden increase of the $T_p/T_e$ ratio at
$\epsilon_p=10^{-1.6}$.  When $\beta_{||p}$= 0.2 the proton and
electron temperatures are in equilibrium ($T_p/T_e=1$) for the
largest $\epsilon_p$ values while for $\beta_{||p}=1$, protons have a factor of 1.2 higher temperature than electrons.

$T_p$ is known to be a strong function of the solar wind speed \citep[e.g.][]{burlaga1973solar,richardson2003radial}, which may affect the observed temperature variations in the ($\epsilon_p, \beta_{||p}$) space. To investigate this speed dependence the $T_p$ data was binned in the $(\epsilon_p,V_{SW})$ space and cross sections were taken along three solar wind speed intervals. The results are shown in Figure \ref{fig:4} where each line was normalized to its peak $T_p$ value. The temperature variations as a function of $\epsilon_p$ are consistent with Figures \ref{fig:2}-\ref{fig:3} and show a sudden enhancement at approximately $\epsilon_p = 10^{-1.6}$ indicating that $T_p$ does have a dependence on $\epsilon_p$ in addition to the dependence on $V_{SW}$.

We note that $\epsilon_p \propto 1/ \sqrt{T_{\perp}}$ and $\beta_{||p} \propto T_{||} $. If only this intrinsic dependence of the variables was
significant we would expect the highest $T_p$ at the lowest $\epsilon_p$, and for a fixed $\beta_{||p}$, $T_{\perp}/T_{||}$ would
decrease as a function of $\epsilon_p$. Neither of these tendencies are observed in Figures \ref{fig:2} or \ref{fig:3}, implying that the amplitude of the turbulent fluctuations is the primary driver of the magnitude of $\epsilon_p$.

\section{Conclusion} \label{sec:conclusion}

In this Letter, we have provided the first statistical test for the
presence of stochastic ion heating of the type predicted by
\citet[]{chandran2010perpendicular}. Our findings are consistent with
the prediction that stochastic heating becomes effective once
gyroscale velocity fluctuations surpass a critical amplitude leading
to perpendicular proton heating. 
We found that the critical
$\epsilon_p$ value in our study is in the range of 0.019 and 0.025 and that
76\% of the studied intervals had an $\epsilon_p$ value larger than
0.025, consistent with stochastic ion heating operating nearly
continuously in the solar wind at 1 AU. Based on the distribution of
the temperature data in the ($\epsilon_p$, $\beta_{||p}$) space, we
make the following conclusions:
\begin{enumerate}
\item If $\beta_{||p}$ = 0.2 and $\epsilon_p \ll \epsilon_{\rm crit}$
  the lowest scalar proton temperatures ($\sim 5\cdot 10^4$ K) were
  measured. The majority of the turbulent energy is absorbed by
  electrons as shown by the low $T_p/T_e$ ratios observed for this
  case.
\item If $\beta_{||p}$ = 0.2 and $\epsilon_p \gtrsim \epsilon_{\rm
  crit}$, an increase in the perpendicular proton temperature was
  identified, with $T_{\perp}/T_{||} \sim 1.20$ for the largest values
  of $\epsilon_p$. The scalar proton temperature increased by a factor
  of 3 compared to the $\beta_{||p}$ = 0.2, $\epsilon_p \ll
  \epsilon_{\rm crit}$ case.
\item If $\beta_{||p}$ = 1 and $\epsilon_p \ll \epsilon_{\rm crit}$,
  no preferential perpendicular heating was observed
  ($T_{\perp}/T_{||} \sim$ 0.96), consistent with non-stochastic
  heating from AW/KAW turbulence.
\item If $\beta_{||p}$ = 1 and $\epsilon_p \gtrsim \epsilon_{\rm
  crit}$, no preferential increase in $T_\perp$ was identified
  ($T_{\perp}/$$T_{||} \sim$ 1.01) and the scalar proton temperature
  reached $1.58 \cdot 10^5$ K, a factor of 3 increase compared to the
  $\beta_{||p}$ = 1, $\epsilon_p \ll \epsilon_{\rm crit}$ case.
\end{enumerate}

The findings above qualitatively agree with the predictions of
\citet[]{chandran2010perpendicular}, which is the main result of this
Letter. We do note that our observed value of $\epsilon_{\rm crit}$ is
an order of magnitude smaller than that reported by
\citet[]{chandran2010perpendicular}, which arises from a prediction
for when more than half of the cascade power near $k_\perp \rho_p =1$
is absorbed by stochastic heating; see their Equations 25, 30, and
31. Their calculation depends sensitively on several dimensionless
parameters characterizing the turbulent fluctuations. Variation in
these parameters may be sufficient to explain the discrepancy in the
value of $\epsilon_{\rm crit}$.

Another potential explanation for this discrepancy arises from \cite{Kasper:2017} suggesting that a majority of preferential minor ion heating occurs within a zone some tens of solar radii from the Sun's surface. It is plausible that the same mechanism preferentially heating the minor ions also heats the protons and a significant fraction of the energy transfer occurs within the preferential heating zone. Thus, the $\epsilon_{\rm crit}$ observed at $1$ AU may not be the actual threshold for the onset of stochastic heating, but rather a value to which $\epsilon_{\rm crit}$ has decayed. Similarly, the observed correlation between $\epsilon_p$ and $T$ may be a remnant of heating closer to the Sun, with plasma which underwent stochastic heating and retained relatively high values of temperature and $\epsilon_p$ compared to other plasma measured at 1 AU. 

\acknowledgments

K.G. Klein was supported by NASA grant NNX16AM23G. J.C. Kasper was supported by NASA Grant NNX14AR78G. Data were sourced from CDAWeb (http://cdaweb.gsfc.nasa.gov/).

\end{document}